\documentclass[12pt]{article}

\usepackage[centertags]{amsmath}
\usepackage{amsmath}
\usepackage{amsfonts}
\usepackage{amssymb}
\usepackage{amsthm}
\usepackage{newlfont}
\usepackage{amscd}
\usepackage{empheq}
\usepackage{epsfig,graphics,graphicx,amsfonts,psfrag}
\usepackage{verbatim}
\usepackage{eucal}

\newcommand{\tr}{\mbox{tr}}
\newcommand{\NN}{{\mathbb N}}

\newcommand{\RR}{{\mathbb R}}
\newcommand{\ZZ}{{\mathbb Z}}

\newcommand{\CC}{{\mathbb C}}
\newcommand{\beq}{\begin{equation}}
\newcommand{\eeq}{\end{equation}}
\newcommand{\ba}{\begin{array}}
\newcommand{\ea}{\end{array}}
\newcommand{\bea}{\begin{eqnarray}}
\newcommand{\eea}{\end{eqnarray}}
\newcommand{\eps}{{\epsilon}}

\usepackage{graphicx}
\usepackage[usenames,dvipsnames]{color}
\usepackage{transparent}
\usepackage{float}

\newtheorem{remark}{Remark}

\begin{document}

\begin{center}
{\bf The linear and nonlinear instability of the Akhmediev breather}

\vskip 10pt

{\it P. G. Grinevich $^{1,3,}$\footnote{The author was supported by the Russian Science Foundation grant 18-11-00316} and P. M. Santini $^{2,4}$}

\vskip 10pt

{\it 
  $^1$ Steklov Mathematical Institute of Russian Academy of Sciences, 8 Gubkina St., Moscow, 199911, Russia,  and 
 L.D. Landau Institute for Theoretical Physics, pr. Akademika Semenova 1a,
Chernogolovka, 142432, Russia

\smallskip

$^2$ Dipartimento di Fisica, Universit\`a di Roma "La Sapienza", and 
Istituto Nazionale di Fisica Nucleare (INFN), Sezione di Roma, 
Piazz.le Aldo Moro 2, I-00185 Roma, Italy}

\vskip 10pt

$^{3}$e-mail:  {\tt pgg@landau.ac.ru}\\
$^{4}$e-mail:  {\tt paolo.santini@roma1.infn.it, paolomaria.santini@uniroma1.it}
\vskip 10pt

{\today}

\end{center}

\begin{abstract}
  The Akhmediev breather (AB) and its M-soliton generalization, hereafter called $AB_M$, are exact solutions of the focusing NLS equation periodic in space and exponentially localized in time over the constant unstable background; they describe the appearance of $M$ unstable nonlinear modes and their interaction, and they are expected to play a relevant role in the theory of periodic anomalous (rogue) waves (AWs) in nature. It is therefore important to establish the stability properties of these solutions under perturbations, to understand if they appear in nature, and in which form. It is rather well established that they are unstable with respect to small perturbations of the NLS equation. Concerning perturbations of these solutions within the NLS dynamics, there is the following common believe in the literature. Let the NLS background be unstable with respect to the first $N$ modes; then i) if the $M$ unstable modes of the $AB_M$ solution are strictly contained in this set ($M<N$), then the $AB_M$ is unstable; ii) if they coincide with this set ($M=N$), the so-called ``saturation of the instability'', then the $AB_M$ solution is neutrally stable. In this paper we argue instead that the $AB_M$ solution is always unstable, even in the saturation case $M=N$, and we prove it in the simplest case $M=N=1$. We first prove the linear instability, constructing two examples of $x$-periodic solutions of the linearized theory growing exponentially in time. Then we investigate the nonlinear instability using some aspects of the recently developed finite gap theory for NLS AWs, showing that i) a perturbed AB initial condition evolves into an exact Fermi-Pasta-Ulam-Tsingou (FPUT) recurrence of ABs described in terms of elementary functions of the initial data, to leading order; ii) the AB solution is more unstable than the background solution, and its instability increases as $T\to 0$, where $T$ is the AB appearance time parameter. Although the AB solution is linearly and nonlinearly unstable, it is relevant in nature, since its instability generates a FPUT recurrence of ABs. These results suitably generalize to the case $M=N>1$.          
\end{abstract}

\section{Introduction}

The self-focusing Nonlinear Schr\"odinger (NLS) equation
\beq\label{eq:nls}
i u_t +u_{xx}+2 |u|^2 u=0, \ \ u=u(x,t)\in\CC, 
\eeq
is the simplest universal model in the description of the propagation of a quasi monochromatic wave in a weakly nonlinear medium; in particular, it is relevant in water waves \cite{Zakharov,AS}, in nonlinear optics \cite{Solli,Bortolozzo,PMContiADelRe}, in Langmuir waves in a plasma \cite{Malomed}, and in the theory of Bose-Einstein condensates \cite{Bludov,Pita}. Its homogeneous solution
\beq\label{back}
u_0(x,t)=\exp(2it), 
\eeq
describing Stokes waves \cite{Stokes} in a water wave context, a state of constant light intensity in nonlinear optics, and a state of constant boson density in a Bose-Einstein condensate, is linearly unstable under monochromatic perturbations of wave number $k$, if $0<|k|<2$, implying the existence of exactly $N=\lfloor L/\pi\rfloor$ unstable modes growing exponentially with growth rates
\begin{equation}
\label{eq:sigmas1}
\sigma_j=k_j\sqrt{4-k^2_j}\,, \ \ \ \ k_j=\frac{2\pi}{L}j, \ \ \ \ 1 \leqslant j \leqslant N,
\end{equation}
the so-called modulation instability (MI) \cite{Talanov,BF,Zakharov}, and this modulation instability (MI) is considered as the main cause for the formation of anomalous (rogue, extreme, freak) waves (AWs) in nature \cite{KharifPeli3,Onorato2}.

The integrable nature of (\ref{eq:nls}) \cite{ZakharovShabat} allows one to construct a large family of exact solutions over the background, including the Peregrine \cite{Peregrine} and Kuznetsov-Ma \cite{Kuznetsov,Ma} solitons, the Akhmediev breather \cite{Akhmed0}, and its elliptic \cite{Akhmed1,Akhmed2} and multi-soliton \cite{ItsRybinSall} generalizations. Concerning the NLS Cauchy problem in which the initial condition consists of a generic small perturbation of the exact background (\ref{back}), what we call the Cauchy problem of the AWs, if such a perturbation is localized, then slowly modulated periodic oscillations described by the elliptic solution of (\ref{eq:nls}) play a relevant role in the longtime regime \cite{Biondini1,Biondini2}. If the initial perturbation is periodic, numerical and real experiments indicate that the solutions of NLS exhibit instead time recurrence \cite{Yuen1,Lake,Yuen3,Akhmed3,Simaeys,Kimmoun,Mussot,PieranFZMAGSCDR}, as well as numerically induced chaos \cite{AblowHerbst,AblowSchobHerbst,AblowHHShober}, in which the almost homoclinic solutions of Akhmediev type seem to play a relevant role \cite{Ercolani,FL,CaliniEMcShober,CaliniShober1,CaliniShober2}. Peregrine and Akhmediev solitons have been observed in experiments; see, for instance: \cite{CHA_observP,KFFMDGA_observP,Yuen3,Kimmoun,Mussot,PieranFZMAGSCDR}.

The Akhmediev breather (AB) and its M-soliton generalization, hereafter called $AB_M$, are exact solutions of the focusing NLS equation periodic in space and exponentially localized in time over the constant unstable background; they describe the appearance of $M$ unstable modes and their nonlinear interaction, and they are expected to play a relevant role in the theory of periodic anomalous (rogue) waves (AWs) in nature. It is therefore important to establish the stability properties of these solutions under perturbations, to understand if they appear in nature, and in which form.

It is well established that they are unstable with respect to small perturbations of the NLS equation \cite{Kimmoun,Soto,CGS,CS1}; see also a finite-gap model describing the numerical instabilities of the AB \cite{GS4}.

General considerations on the perturbation theory within the NLS dynamics were made in \cite{LiMcL}. Concerning perturbations of the $AB_M$ solutions within the NLS dynamics,  
the problem was first posed and investigated in \cite{CaliniShober2,CaliniShober3,CaliniShober4}, generating the following common believe in the literature. Let the NLS background be unstable with respect to the first $N$ modes; then i) if the $M$ unstable modes of the $AB_M$ solution are strictly contained in this set ($M<N$), the $AB_M$ is linearly unstable; ii) if they coincide with this set ($M=N$), the so-called ``saturation of the instability'', then the $AB_M$ solution is linearly and neutrally stable. The conclusion about the neutral stability was based on the following argument: in a generic situation, an arbitrary solution of the linearized equation admits an expansion in terms of the $x$-periodic squared eigenfunctions, and  for the expansion about the AB type solutions with $M=N$, all the squared eigenfunctions are bounded in $t$.

While the instability of the $AB_M$ solution if $M<N$ was correctly established in \cite{CaliniShober2,CaliniShober3,CaliniShober4}, the neutral stability of the saturated case $M=N$ is not true. As we shall see in this paper: the $AB_M$ solution is always unstable, even if $M=N$, although this case is less unstable than the case $M<N$. The reason is that, for the $AB_M$ solutions, the situation is non-generic: the spectral curve has double points (see \S 2), therefore the squared eigenfunction decomposition includes also some special combination of derivatives of the squared eigenfunctions with respect to the spectral parameter, and these additional terms grow exponentially in time. 

The paper is organized in the following way.

In \S 2 we present two qualitative, but convincing arguments for the linear and nonlinear instability of the $AB_M$ solution, and in \S 3 we argue that the $AB_M$ solution is always linearly unstable, even in the saturation case $M=N$, and we prove it in the simplest case $M=N=1$ constructing periodic solutions of the linearized theory growing exponentially in time. We construct these ``missing'' periodic solutions of the linearized equation by direct matching of the coefficients. We plan to provide a regular procedure for constructing the squared eigenfunction decomposition for non-generic situations in a subsequent paper.

In addition, motivated by the recent solution, to leading order and in terms of elementary functions, of the periodic Cauchy problem of NLS AWs:
\beq\label{eq:nls_cauchy1}
\ba{l}
u(x,0)=1+\varepsilon v(x), \ 0<\varepsilon\ll 1, \ v(x+L)=v(x), \\
v(x)=\sum_{j=1}^{\infty}(c_j e^{i k_j x}+c_{-j}e^{-i k_j x}),\ \ k_j=\frac{2\pi}{L}j,
\ea
\eeq
in the case of a finite number $N$ of unstable modes \cite{GS1,GS2}, in \S 4 we show, in the simplest case $N=M=1$,  how this linear instability  develops into the full nonlinear regime, evolving into an AW recurrence (of Fermi-Pasta-Ulam-Tsingou (FPUT) type) of ABs, thus establishing the nonlinear instability of the AB and, at the same time, its relevance in natural phenomena! We also show that, not only the AB solution is unstable in the case $M=N=1$, but it is more unstable than the background solution with respect to the same periodic perturbations! \S 5 is dedicated to conclusions and remarks.\\

The solution \cite{GS1,GS2} of the NLS periodic Cauchy problem \eqref{eq:nls_cauchy1} of the AWs is based on the proper adaptation of the  finite-gap method \cite{Novikov,Dubrovin,ItsMatveev,Lax,MKVM,Krichever} (see \cite{ItsKotlj} for its first application to NLS), to it. See also \cite{GS3} for an alternative approach to the study of the AW recurrence, based on matched asymptotic expansions; see \cite{GS5} for the analytic study of the phase resonances in the AW recurrence; see \cite{San}, \cite{CS2} and \cite{CS3} for the analytic study of the AW recurrence in other NLS type models: respectively the PT-symmetric NLS equation \cite{AM1}, the Ablowitz-Ladik model \cite{AL}, and the massive Thirring model \cite{Thirring,Mikhailov}.

The zero-curvature representation of (\ref{eq:nls}) is \cite{ZakharovShabat}:
\begin{equation}
\label{eq:lp-x}
\vec\psi_x(\lambda,x,t)=\left[-i\lambda \sigma_3 +iU(x,t)\right]\vec\psi(\lambda,x,t),
\end{equation}
\begin{equation}
\label{eq:lp-t}
\vec\psi_t(\lambda,x,t)=\hat V(\lambda,x,t)\vec\psi(\lambda,x,t),
\end{equation}
\beq\label{def_U}
\sigma_3=\begin{bmatrix}  1 & 0 \\ 0 & -1 \end{bmatrix}, \ U = \begin{bmatrix}  0 & u\\ \bar u & 0  \end{bmatrix}, \
\vec\Psi= \left [\begin {array}{c} \Psi_1 \\ \Psi_2 \end {array}\right ],
\eeq
$$
\hat V= \left[\begin {array}{cc} -2 i \lambda^2 + i u\overline{u} & 2 i \lambda u - u_x
\\\noalign{\medskip} 2 i \lambda \overline{u} +\overline{u_x} & 2 i \lambda^2- i u\overline{u} 
\end {array}
\right ].
$$
The fundamental matrix solution $\hat T(\lambda,x,y,t)$ of (\ref{eq:lp-x}),\eqref{eq:lp-t},\eqref{def_U} in the $x$-periodic problem, such that $\hat T(\lambda,y,y,t)=E$, where $E$ is the identity matrix (see \cite{Faddeev}), is an entire function of $\lambda$. The eigenvalues and eigenvectors of the monodromy matrix $T(\lambda,t)=\hat T(\lambda,L,0,t) $ are defined on a two-sheeted covering of the $\lambda$-plane. This Riemann surface $\Gamma$ is called the spectral curve and does not depend on time.  The eigenvectors of $T(\lambda,t)$ are the Bloch eigenfunctions
\beq
\label{eq:bloch1}
\vec\Psi(\gamma,x+L,t) =e^{iLp(\gamma)} \vec\Psi(\gamma,x,t), \ \ \gamma \in \Gamma.
\eeq

The main spectrum is exactly the projection of the set $\{\gamma\in\Gamma$, Im $p(\gamma)=0 \}$ to the $\lambda$-plane, and it is a constant of motion with respect to the NLS evolution. The end points of the spectrum are the branch points and the double points of $\Gamma$, at which $e^{iLp(\gamma)}=\pm 1$, or, equivalently, $\tr T(\lambda)=\pm 2$.

For the background (\ref{back})  the curve $\Gamma_0$ is rational and defined by equation: $\mu^2=\lambda^2+1$. The corresponding monodromy matrix: $\tr T_0(\lambda) = 2 \cos (\mu L)$ defines the branch points $(\lambda^{\pm}_0,\mu_0)=(\pm i,0)$ and the double points $(\lambda^{\pm}_n,\mu_n)=(\pm\sqrt{(n\pi/L)^2-|a|^2},n\pi/L)$, $n\in\ZZ,~n\ne 0$. 

Under the perturbation (\ref{eq:nls_cauchy1}) of the background, the branch points $\lambda^{\pm}_0=\pm i$ become $E_0= i+O(\eps^2)$ and $\bar E_0$ (hereafter $\bar f$ is the complex conjugate of $f$), and all double points $\lambda^{\pm}_n$, $n\ge1$, generically split into pairs of square root branch points, generating infinitely many gaps. If $1\le n\le N$, where $N$ is the number of unstable modes, $\lambda^+_n$ and $\lambda^-_n$ split into the pair of branch points ($E_{2n-1},E_{2n}$) and ($\bar E_{2n-1},\bar E_{2n}$) respectively, where
\beq\label{eq:bps2}
\ba{l}
E_{l}=\lambda_n\mp\frac{1}{2\lambda_n}\sqrt{\alpha_n\beta_n}+O(\epsilon^2), \ \ l=2n-1,2n, 
\ea
\eeq
and
\beq\label{def_alpha_beta}
\ba{l}
\alpha_n :=\eps\left(e^{-i\phi_n}\overline{c_n}-e^{i\phi_n}c_{-n}\right),\quad \beta_n :=\eps\left(e^{i\phi_n}\overline{c_{-n}}-e^{-i\phi_n}c_n\right),\\[2mm]
\phi_n:=\arccos\left(k_n/2\right).
\ea
\eeq
In \cite{GS1,GS2} the solution of the generic AW Cauchy problem was approximated by a finite-gap one obtained closing all gaps near the real line, since they correspond to the stable modes and contribute to the solution to $O(\eps)$.

If $\pi< L < 2\pi$, then $N=1$ and the solution is well approximated by a genus 2 exact solution on a Riemann surface with $O(\eps)$ handles, expressible, to leading order, in terms of elementary functions of the initial data. Then the solution of the Cauchy problem to leading order (up to $O(\epsilon)$ corrections), in the finite interval $0\le t \le T_0$,  reads as follows \cite{GS1,GS3}:
\begin{align}\label{unif_sol_Cauchy_1}
u(x,t)&=\sum\limits_{m=0}^n {\cal A}\Big(x,t;\phi_1,x^{(m)},t^{(m)} \Big)
e^{i\rho^{(m)}}- \\
 &-\frac{1-e^{4in\phi_1}}{1-e^{4i\phi_1}} e^{2it}, \ \ x\in [0,L], \nonumber
\end{align}
where the parameters $x^{(m)},~ t^{(m)},~\rho^{(m)},~m\ge 0$, are defined as:
\beq\label{parameters_1n}
\ba{l}
x^{(m)}=X^{(1)}+(m-1)\Delta X, \ \ t^{(m)}=T^{(1)} + (m-1)\Delta T, \\
X^{(1)}=\frac{\arg\alpha_1}{k_1} +\frac{L}{4}, \ \ \Delta X =\frac{\arg(\alpha_1\beta_1)}{k_1}, \ \ (\!\!\!\!\!\mod L),\\
T^{(1)}\equiv \frac{1}{\sigma_1}\log\left(\frac{\sigma^2_1}{2|\alpha_1|} \right), \ \
\Delta T = \frac{1}{\sigma_1} \log\left(\frac{\sigma^4_1}{4|\alpha_1\beta_1|}\right), \,  \\
\rho^{(m)}=2\phi_1+(m-1)4\phi_1 ,\ \
n = \left\lceil \frac{T - T^{(1)}}{\Delta T} +\frac{1}{2} \right\rceil ,
\ea
\eeq
and function~${\cal A}$ is the AB:
\begin{equation}
\label{eq:akh1}
\begin{gathered}
{\cal A}(x,t;\theta,X,T):=e^{2it}\frac{\cosh[\sigma(\theta)(t-T)+2i\theta]+
\sin\theta\cos[k(\theta)(x-X)]}{\cosh[\sigma(\theta)(t-T)]-
\sin\theta\cos[k(\theta)(x-X)]}\,,
\\[2mm]
k(\theta)=2\cos\theta, \ 
\sigma(\theta)=k(\theta)\sqrt{4-k^2(\theta)}=2\sin(2\theta),
\end{gathered}
\end{equation}
exact solution of NLS for all real parameters $\theta,X,T$.

The solution (\ref{unif_sol_Cauchy_1})-(\ref{eq:akh1}) shows an exact recurrence of AWs described by the AB, whose parameters change at each appearance according to (\ref{parameters_1n}). It provides the quantitative theory of the FPUT \cite{FPU} type recurrence, without thermalisation, of the NLS AWs, in terms of elementary functions. $X^{(1)}$ and $T^{(1)}$ are respectively the position and the time of the first appearance; $\Delta X$ is the $x$-shift of the position of the AW between two consecutive appearances, and $\Delta T$ is the recurrence time (the time between two consecutive appearances). Therefore $T^{(1)}$ and $\Delta T$ are the characteristic times of the AW recurrence.

We remark that formulas (\ref{unif_sol_Cauchy_1})-(\ref{eq:akh1}), in perfect quantitative agreement with the output of the corresponding numerical experiments \cite{GS1}, were successfully tested in nonlinear optics experiments involving a photorefractive crystal \cite{PieranFZMAGSCDR}.

\section{Two qualitative arguments for the instability of the $AB_M$}

We begin with two qualitative, but convincing arguments for the instability of the $AB_M$ even in the saturation case. Both arguments are based on the well-known fact that the $AB_M$ solution reduces, for large $|t|$ and up to phase constants, to the background solution (\ref{back}). \\[2mm]
1) {\bf Linear instability argument}. The NLS equation linearized about the $AB_M$ solution describes the linear stability properties of the $AB_M$. Since, for large $|t|$, the $AB_M$ reduces to the background (\ref{back}), the linearized equation about the $AB_M$ tends, for large $|t|$, to the linearized equation about the background, whose solutions generically blow up exponentially in time. Therefore such a blow up is expected to be present, asymptotically, also in the linearized equation about the $AB_M$, implying its linear instability.\\[2mm]
2) {\bf Nonlinear instability argument}. Since, for large $|t|$, the $AB_M$ reduces to the background, the $AB_M$ shares with the background solution the main spectrum of the spectral problem \eqref{eq:lp-x} for periodic potentials, a constant of motion for the NLS dynamics, consisting of the square root branch points $\pm i$ and of the infinitely many double points $\lambda^{\pm}_n=\pm\sqrt{(n\pi/L)^2-1}$, $n\in\ZZ,~n\ne 0$. It follows that a generic perturbation of the $AB_M$ initial condition will generically resolve such a degeneration opening infinitely many small gaps, and the $2N$ gaps associated with the unstable modes will generate $O(1)$ deviations from the $AB_M$ dynamics, implying the nonlinear instability of the $AB_M$ solution, even when $M=N$.

\section{Linear instability of the AB solution}

In this section we prove the linear instability of the AB constructing examples of periodic solutions of the NLS equation linearized about the AB growing exponentially in time. To do so, we  find it convenient to remove the phase factor $\exp(2it)$ from the NLS solutions through the transformations
\beq
u(x,t)\to \exp(2it) u(x,t), \ \ \ \ \vec\psi \to \exp(i\sigma_3 t)\vec\psi,
\eeq
dealing now with the NLS equation in the form:
\begin{equation}\label{NLS2}
iu_t + u_{xx} + 2|u|^2 u -2 u =0.
\end{equation}

\subsection{The space of symmetries of the NLS equation}

To establish the linear instability of the AB with respect to small periodic perturbations, it is enough to construct examples of $x$-periodic and exponentially growing in time solutions $w(x,t)$ of the focusing NLS equation \eqref{NLS2} linearized about the AB solution, the so-called ``linearized NLS equation'':
\beq\label{linearizedNLS}
i w_t+w_{xx}+4 |u(x,t)|^2 w +2 u^2(x,t) \bar w -2 w =0, \ \ w=w(x,t)\in\CC ,
\eeq
where $u(x,t)$ is the AB solution of \eqref{NLS2}.

It is well-known that the solutions of (\ref{linearizedNLS}) are (infinitesimal generators of the) symmetries of the NLS equation \eqref{NLS2} (see, f.i., \cite{FF}). It is also well-known \cite{AKNS} that the so-called squared eigenfunctions of the spectral problem of NLS are the ``generators'' of solutions of (\ref{linearizedNLS}) in the following way.

Let $\vec\psi(\lambda,x,t)=(\psi_1(\lambda,x,t),\psi_2(\lambda,x,t))^T$ and $\vec\varphi(\lambda,x,t)=(\varphi_1(\lambda,x,t),\varphi_2(\lambda,x,t))^T$ be two eigenfunctions of the zero curvature representation of NLS \eqref{NLS2}. Then the following bilinear forms
\beq\label{sym1}
\ba{l}
\left<\vec\psi(\lambda,x,t),\vec\varphi(\lambda,x,t)\right>_+:=\psi_1(\lambda,x,t)\varphi_1(\lambda,x,t)+\overline{\psi_2(\lambda,x,t)\varphi_2(\lambda,x,t)},\\
\left<\vec\psi(\lambda,x,t),\vec\varphi(\lambda,x,t)\right>_-:=i\left[\psi_1(\lambda,x,t)\varphi_1(\lambda,x,t)-\overline{\psi_2(\lambda,x,t)\varphi_2(\lambda,x,t)}\right]
\ea
\eeq
are solutions of (\ref{linearizedNLS}). We remark that $\left<\cdot,\cdot\right>_-$ can be obtained from $\left<\cdot,\cdot\right>_+$ through the transformation $\vec\psi\to i\vec\psi$, since $i\vec\psi$ is eigenfunction if $\vec\psi$ is eigenfunction. In the following we find it convenient to use the Hadamard product of two vectors $\vec a=(a_1,a_2)^T$ and $\vec b=(b_1,b_2)^T$: 
\beq
\vec a \circ \vec b := \begin{bmatrix} a_1 b_1 \\ a_2 b_2 \end{bmatrix},
\eeq
in terms of which the two symmetries \eqref{sym1} can be written as
\beq
\ba{l}
\left<\vec\psi,\vec\varphi\right>_+ =\left(\vec\psi \circ \vec\varphi\right)_1+\overline{\left(\vec\psi \circ \vec\varphi\right)_2},\\[2mm]
\left<\vec\psi,\vec\varphi\right>_- =i\left[\left(\vec\psi \circ \vec\varphi\right)_1-\overline{\left(\vec\psi \circ \vec\varphi\right)_2}\right].
\ea
\eeq

To construct the eigenfunctions corresponding to the AB solution of \eqref{NLS2}, describing the instability of the first mode $k_1=2\pi/L$, we use the classical Darboux transformation (DT) \cite{Matveev0} from the pair $(u_0,\vec\psi_0)$ corresponding the background solution $u_0=1$ of \eqref{NLS2} to the pair $(u,\vec\psi)$ corresponding to the AB solution of \eqref{NLS2}.\\
Consider the two independent eigenfunctions corresponding to the background $u_0=1$ of \eqref{NLS2}:
\beq\label{def_psi_0}
\ba{l}
\vec\psi^{\pm}_0(\lambda,x,t)=\left[
  \ba{c}
  \sqrt{\mu\mp\lambda} \\
  \pm\sqrt{\mu\pm\lambda}
  \ea
\right]e^{\pm\theta}, \\
\theta=i\mu x+2i\mu\lambda t ,
  \ea
  \eeq
where
  \beq\label{lambda_mu}
\mu^2=\lambda^2 +1.
\eeq
From \eqref{def_psi_0} and \eqref{lambda_mu} it follows that, to construct solutions periodic in $x$ and hyperbolic in $t$, we have to assume that
\beq
\mu\in\RR, \ \ \ \ \lambda\in i\RR, \ |\lambda |\le 1.
\eeq

Let $\vec q(\lambda)$ and $\vec r(\lambda)$ be respectively the sum and difference of the eigenfunctions \eqref{def_psi_0} (hereafter we omit the $x,t$ dependence when it is not necessary):
\begin{equation}
\begin{split}
&\vec q(\lambda) =  \begin{bmatrix}  q_1(\lambda)  \\ q_2(\lambda) \end{bmatrix} =
\begin{bmatrix}  \sqrt{\mu-\lambda}\, e^{\theta(\lambda)}  +   \sqrt{\mu+\lambda}\, e^{-\theta(\lambda)}  \\  \sqrt{\mu+\lambda}\,e^{\theta(\lambda)}   -  \sqrt{\mu-\lambda}\, e^{-\theta(\lambda)}  \end{bmatrix}, \\
&\vec r(\lambda) =  \begin{bmatrix}  r_1(\lambda)  \\ r_2(\lambda) \end{bmatrix} =
\begin{bmatrix}  \sqrt{\mu-\lambda}\,e^{\theta(\lambda)}   -   \sqrt{\mu+\lambda}\, e^{-\theta(\lambda)}  \\  \sqrt{\mu+\lambda}\,e^{\theta(\lambda)}   +  \sqrt{\mu-\lambda}\, e^{-\theta(\lambda)}  \end{bmatrix},
\end{split}
\end{equation}
and let
  \beq
\vec q=\left[
    \ba{c}
  q_1 \\
  q_2
  \ea
  \right]=\vec q(\lambda_1), \ \ \vec r=\left[
    \ba{c}
  r_1 \\
  r_2
  \ea
  \right]=\vec r(\lambda_1),
  \eeq
  where
  \beq
  \ba{l}
  \lambda_1=\sqrt{\mu^2_1 -1}, \ \ \ \mu_1=\frac{k_1}{2}=\frac{\pi}{L} \\
   \ea
   \eeq
   is the first and only unstable mode (assuming hereafter that $\pi <L<2\pi$), and
   \beq
k=k_1=2\mu_1, \ \  \sigma =\sigma_1= -4 i \lambda_1\mu_1, \ \ \theta(\lambda_1)=\frac{1}{2}\left(ikx-\sigma t \right).
   \eeq

  Introduce the Darboux matrix (see, f.i., \cite{Matveev0,Yurov})
  \beq\label{Darboux}
  \ba{l}
  \hat D(\lambda)=(\lambda-\lambda_1)E +\frac{2\lambda_1}{|q_1|^2 +|q_2|^2}\left[
    \ba{c}
  -\overline{q_2} \\
  \overline{q_1}
  \ea
  \right][-q_2,q_1],
\ea
\eeq
then the DT
\beq\label{DT}
\vec\psi(\lambda)=\hat D(\lambda)\vec\psi_0(\lambda)
\eeq
maps any eigenfunction $\vec\psi_0(\lambda)$ corresponding to the background $u_0=1$ of \eqref{NLS2} to an eigenfunction $\vec\psi(\lambda,x,t)$ corresponding to the solution
\beq\label{AB1}
u(x,t)=1-\frac{4\lambda_1 q_1\overline{q_2}}{Den}= \frac{(\lambda_1^2+\mu_1^2)\cosh( \sigma t ) + i \lambda_1 \sin(k x) + 2\mu_1\lambda_1 \sinh(\sigma t)  }{\cosh( \sigma t ) - i \lambda_1 \sin(k x)},
\eeq
of \eqref{NLS2}, reducing to the AB solution (\ref{eq:akh1}) after multiplying by $\exp(2it)$ and using the $x$ and $t$ translation symmetries, where
\beq
Den(x,t)=|q_1 |^2+|q_2 |^2=4\left[\cosh(\sigma t)-i\lambda_1\sin(kx) \right]=\frac{2}{k}\left[2k\cosh(\sigma t)+\sigma\sin(kx) \right].
\eeq

It is well-known that, for $\lambda=\lambda_1$, the range of the Darboux matrix (\ref{Darboux}) has dimension $1$, and the DT cannot generate two independent eigenfunctions. In particular, applying the DT to $\vec q(\lambda),\vec r(\lambda)$: 
\beq\label{def_chi_1}
\ba{l}
\vec\chi_+(\lambda)=\hat D(\lambda)\vec q(\lambda), \\
\vec\chi_-(\lambda)=\hat D(\lambda)\vec r(\lambda),
\ea
\eeq
one obtains two independent eigenfunctions corresponding to the AB (\ref{AB1}) if $\lambda\neq\lambda_1$. If $\lambda=\lambda_1$, $\vec\chi_+$ becomes zero, and the two symmetries \eqref{sym1}, constructed using only the eigenfunction $\chi^{(0)}_-:=\vec\chi_-(\lambda_1)$, read
\beq
\ba{l}
\left<\chi^{(0)}_-,\chi^{(0)}_-\right>_+=16\lambda_1^2 \mu_1^2~\frac{\cos(k x)\left[\mu_1\cosh(\sigma t)+\lambda_1 \sinh(\sigma t)\right]}{[\cosh(\sigma t)-i\lambda_1 \sin(k x)]^2}, \\
\ \\
\left<\chi^{(0)}_-,\chi^{(0)}_-\right>_-=-16i\lambda_1^2 \mu_1^2~\frac{1-i \sin(k x)\left[\mu_1\sinh(\sigma t)+\lambda_1 \cosh(\sigma t) \right]}{[\cosh(\sigma t)-i\lambda_1 \sin(k x)]^2}.
\ea
\eeq
They are $x$-periodic and exponentially localized in time, and do not describe any blow up. This is presumably the reason why it was argued in \cite{CaliniShober2,CaliniShober3} that the AB is neutrally stable if $N=M=1$. But the non generic spectral nature of the AB, whose  spectral curve has infinitely many double points, implies that the corresponding squared eigenfunction decomposition includes also some special combinations of derivatives of the squared eigenfunctions with respect to the spectral parameter, and, as we shall see, these additional terms grow exponentially in time. 

The following remarks are relevant at this point. \\
1) The space of solutions of (\ref{linearizedNLS}) is a vector space over $\RR$ (if $w_1$ and $w_2$ are solutions of (\ref{linearizedNLS}), also their linear combination with real constant coefficients is solution).\\
2) The bilinear forms \eqref{sym1} are solutions of (\ref{linearizedNLS}) $\forall \lambda\in\CC$. \\
3) If $w(\lambda,x,t)$ is a solution of (\ref{linearizedNLS}) $\forall\lambda\in\CC$, also $\left(\partial_{\lambda}+\partial_{\overline{\lambda}}\right)^nw(\lambda,x,t)$ and $\left(i\left(\partial_{\lambda}-\partial_{\overline{\lambda}}\right)\right)^nw(\lambda,x,t)$ are solutions of (\ref{linearizedNLS}), for $n\in\NN^+$ (or the analogous formulas obtained replacing $\partial_{\lambda}$ by $\partial_{\mu}$).

The third property, consequence of the first two, is what we use in the following to construct the wanted solutions. Since we deal with functions of $\lambda$ and $\mu$, related by (\ref{lambda_mu}), it is convenient to introduce the total derivative
\beq
D_{\mu}=\partial_{\mu}+\frac{\partial\lambda}{\partial\mu}\partial_{\lambda}=\partial_{\mu}+\frac{\mu}{\lambda}\partial_{\lambda};
\eeq
then remark 3) implies that
\beq
\ba{l}
\left(D_{\mu}+D_{\bar\mu} \right)^n \left<\vec\psi(\lambda),\vec\varphi(\lambda)\right>_+=D^n_{\mu}\left(\psi_1(\lambda)\varphi_1(\lambda) \right)+\overline{D^n_{\mu}\left(\psi_2(\lambda)\varphi_2(\lambda) \right)}, \\
\left(D_{\mu}+D_{\bar\mu} \right)^n \left<\vec\psi(\lambda),\vec\varphi(\lambda)\right>_-=i\left[D^n_{\mu}\left(\psi_1(\lambda)\varphi_1(\lambda) \right)-\overline{D^n_{\mu}\left(\psi_2(\lambda)\varphi_2(\lambda) \right)}\right]
\ea
\eeq
are solutions of (\ref{linearizedNLS}) $\forall\lambda\in\CC$ and $n\in\NN$.

\subsection{Construction of the wanted symmetries}

The goal of this section is to use the last result to construct examples of $x$-periodic symmetries of NLS growing exponentially in $t$ for large $|t|$. In the following we make use of the notation
\beq
f^{(n)}(x,t):=D^n_{\mu}f(\lambda,x,t)|_{\lambda=\lambda_1}, \ \ \ n\in\NN.
\eeq

One can show that
\beq
\ba{l}
D_{\mu} \vec q(\lambda) = \frac{1}{2\lambda} \begin{bmatrix}  -r_1(\lambda)  \\ r_2(\lambda) \end{bmatrix} 
+ \theta_{\mu}(\lambda)  \vec r(\lambda), \\
D_{\mu} \vec r(\lambda) = \frac{1}{2\lambda} \begin{bmatrix}  -q_1(\lambda)  \\ q_2(\lambda) \end{bmatrix} 
+  \theta_{\mu}(\lambda) \vec q(\lambda), \\
D^2_{\mu} \vec q(\lambda) =  \bigg[\frac{1}{4\lambda^2} + \theta^2_{\mu}(\lambda)   \bigg] \vec q(\lambda) - \frac{\mu}{2\lambda^3} \begin{bmatrix} - r_1(\lambda)  \\ r_2(\lambda) \end{bmatrix} 
+  \frac{\theta_{\mu}(\lambda)}{\lambda}  \begin{bmatrix}  -q_1(\lambda)  \\ q_2(\lambda) \end{bmatrix} +\theta_{\mu\mu}(\lambda)  \vec r(\lambda).
\ea
\eeq
We rewrite the vectors $\vec\chi_{\pm}$ in \eqref{def_chi_1} in the form
\beq\label{def_chi_2}
\ba{l}
\vec\chi_{+}(\lambda)=(\lambda-\lambda_1)\begin{bmatrix} q_1(\lambda)  \\ q_2(\lambda) \end{bmatrix}+2\lambda_1 \frac{pr_{+}(\lambda)}{|q_1|^2+ |q_2|^2}\begin{bmatrix} -\bar q_2  \\ \bar q_1 \end{bmatrix}, \\
\vec\chi_{-}(\lambda)=(\lambda-\lambda_1)\begin{bmatrix} r_1(\lambda)  \\ r_2(\lambda) \end{bmatrix}+2\lambda_1 \frac{pr_{-}(\lambda)}{|q_1|^2+ |q_2|^2}\begin{bmatrix} -\bar q_2  \\ \bar q_1 \end{bmatrix}, 
\ea
\eeq
where
\beq
pr_{+}(\lambda):=[-q_2,q_1] \begin{bmatrix} q_1(\lambda)  \\ q_2(\lambda) \end{bmatrix}, \ \ pr_-(\lambda) =[-q_2,q_1] \begin{bmatrix} r_1(\lambda)  \\ r_2(\lambda) \end{bmatrix},
\eeq
implying
\begin{equation}
\begin{split}
pr^{(0)}_{-} & = 4\mu_1,\\ 
pr^{(1)}_{-} & = 2\left[1+ \frac{\sinh(\theta) }{\lambda_1}\right],\\
pr^{(0)}_{+} & = 0,\\ 
pr^{(1)}_{+} & = \frac{2 \cosh(\theta)}{\lambda_1} + 4\mu_1\theta_{\mu}=  \frac{2 \cosh(\theta)}{\lambda_1} + pr_{-0}\,\theta_{\mu} ,\\
pr^{(2)}_{+} & = \frac{-2\mu_1 \cosh(\theta)}{\lambda_1^3} +  4\left[1+ \frac{\sinh(\theta) }{\lambda_1}\right] \theta_{\mu} +4\mu_1\theta_{\mu\mu}=\\
& = \frac{-2\mu_1 \cosh(\theta)}{\lambda_1^3} +  2pr^{(1)}_{-}\, \theta_{\mu} + pr^{(0)}_{-}\,\theta_{\mu\mu}   .
\end{split}
\end{equation}
After some (tedius) algebra we obtain:
\begin{equation}
\begin{split}
\tilde\chi^{(0)}_{-} &= \frac{8\lambda_1\mu_1 }{q_1\bar q_1 + q_2\bar q_2}\begin{bmatrix} -\bar q_2  \\ \bar q_1 \end{bmatrix} ,\\
\tilde\chi^{(1)}_{-} &= \frac{\mu_1}{\lambda_1} \begin{bmatrix} r_1  \\ r_2 \end{bmatrix} +  
\frac{4[\lambda_1  + \sinh(\theta)]  }{q_1\bar q_1 + q_2\bar q_2} \begin{bmatrix} -\bar q_2  \\ \bar q_1 \end{bmatrix},\\
\tilde\chi^{(0)}_{+} &= \begin{bmatrix} 0  \\ 0 \end{bmatrix},\\
\tilde\chi^{(1)}_{+} &=  \frac{\mu_1}{\lambda_1} \begin{bmatrix} q_1  \\ q_2 \end{bmatrix} + \frac{4 \cosh(\theta)  } {q_1\bar q_1 + q_2\bar q_2} \begin{bmatrix} -\bar q_2  \\ \bar q_1 \end{bmatrix}  + \theta_{\mu}  \tilde\chi^{(0)}_{-}, \\
\tilde\chi^{(2)}_{+} &=-\frac{1}{\lambda_1^3} \begin{bmatrix} q_1  \\ q_2 \end{bmatrix}+ \frac{\mu_1}{\lambda_1^2} \begin{bmatrix} -r_1  \\ r_2 \end{bmatrix} -  \frac{4\mu_1 \cos(k\,x)}{\lambda_1^2 (q_1\bar q_1 + q_2\bar q_2)}\begin{bmatrix} -\bar q_2  \\ \bar q_1 \end{bmatrix} + 2\theta_{\mu} \tilde\chi^{(1)}_{-} + \theta_{\mu\mu} \tilde\chi^{(0)}_{-}.  
\end{split}
\end{equation}

In the construction of the wanted symmetries, it is also useful to evaluate the eigenfunction corresponding to the background at the branch point $\lambda=i$ of its main spectrum. From
\beq
\vec \phi(\lambda) = \frac{1}{\sqrt{\mu-\lambda}}\vec\psi^{+}_0(\lambda,x,t)= \begin{bmatrix}  \phi_1(\lambda)  \\ \phi_2(\lambda) \end{bmatrix} =
\begin{bmatrix}  1  \\  (\mu+\lambda) \end{bmatrix}\,e^{\theta(\lambda)}   
\eeq
it follows that:
\begin{equation}
\begin{split}
&D_{\mu} \vec \phi(\lambda) = \begin{bmatrix}  0  \\ \frac{\mu+\lambda}{\lambda} \end{bmatrix}\,e^{\theta(\lambda)}   
+ \theta_{\mu}(\lambda)  \vec\phi(\lambda), \\
\end{split}
\end{equation}
\begin{equation}
\begin{split}
&\vec \phi^{(0)} =\vec \phi(\lambda)\Big|_{\lambda=i}   = \begin{bmatrix}  1  \\ i  \end{bmatrix},\\
&\vec\phi^{(1)} = D_{\mu} \vec \phi(\lambda)\Big|_{\lambda=i}   = \begin{bmatrix}  0  \\1 \end{bmatrix} 
+ (ix - 2t)  \vec\phi^{(0)}, \\
\end{split}
\end{equation}
and the DT of $\vec \phi(\lambda)$, evaluated at $\lambda=i$, gives
\begin{equation}
\begin{split}
&\tilde\phi^{(0)} = \frac{1} {|q_1|^2+ |q_2|^2}\begin{bmatrix} i (|q_1|^2+ |q_2|^2)  - \lambda_1 (|q_1|^2- |q_2|^2)  - 2i\lambda_1 \bar q_2 q_1 ) \\ 
- (|q_1|^2+ |q_2|^2)  + i \lambda_1 (|q_1|^2- |q_2|^2)  - 2\lambda_1 \bar q_1 q_2) \end{bmatrix}=\\
&= \frac{\mu_1}{\cosh(\sigma t ) - i \lambda_1 \sin(k x) }  \begin{bmatrix} i\mu_1\cosh(\sigma t )   - \lambda_1 \cos(k ,x) + i\lambda_1\sinh(\sigma t) \\ 
-\mu_1\cosh(\sigma t )  + i \lambda_1 \cos(k\,x)  + \lambda_1\sinh(\sigma t)  \end{bmatrix}.
\end{split}
\end{equation}

Consider now the following basic symmetries:
\begin{equation}
\begin{split}
&\left.\left<\chi_{+}(\lambda), \chi_{+}(\lambda)\right>_{\pm} \right|_{\lambda=\lambda_1}=0,\\
&\left.\left<\chi_{+}(\lambda), \chi_{-}(\lambda)\right>_{\pm} \right|_{\lambda=\lambda_1}=0,\\
&\left.\left<\chi_{-}(\lambda), \chi_{-}(\lambda)\right>_{\pm} \right|_{\lambda=\lambda_1}= \left<\chi^{(0)}_{-}, \chi^{(0)}_{-}\right>_{\pm} ,\\
&\left.(D_{\mu}+D_{\bar\mu})\left<\chi_{+}(\lambda), \chi_{+}(\lambda)\right>_{\pm} \right|_{\lambda=\lambda_1}=0,\\
&\left.(D_{\mu}+D_{\bar\mu})\left<\chi_{+}(\lambda), \chi_{-}(\lambda)\right>_{\pm} \right|_{\lambda=\lambda_1}= 
\left<\chi^{(1)}_{+}, \chi^{(0)}_{-}\right>_{\pm} ,\\
&\left.(D_{\mu}+D_{\bar\mu})\left<\chi_{-}(\lambda), \chi_{-}(\lambda)\right>_{\pm} \right|_{\lambda=\lambda_1}= 2\left<\chi^{(1)}_{-}, \chi^{(0)}_{-}\right>_{\pm} ,\\
&\left.(D_{\mu}+D_{\bar\mu})^2\left<\chi_{+}(\lambda), \chi_{+}(\lambda)\right>_{\pm} \right|_{\lambda=\lambda_1}= 2 \left<\chi^{(1)}_{+}, \chi^{(1)}_{+}\right>_{\pm} ,\\
&\left.(D_{\mu}+D_{\bar\mu})^2\left<\chi_{+}(\lambda), \chi_{-}(\lambda)\right>_{\pm} \right|_{\lambda=\lambda_1}= 
\left<\chi^{(2)}_{+}, \chi_{-0}\right>_{\pm} + 2 \left<\chi^{(1)}_{+}, \chi^{(1)}_{-}\right>_{\pm} ,\\
&\left.(D_{\mu}+D_{\bar\mu})^2\left<\chi_{-}(\lambda), \chi_{-}(\lambda)\right>_{\pm} \right|_{\lambda=\lambda_1}= 
2\left<\chi^{(2)}_{-}, \chi^{(0)}_{-}\right>_{\pm} + 2 \left<\chi^{(1)}_{-}, \chi^{(1)}_{-}\right>_{\pm}.
\end{split}
\end{equation}
It turns out that the symmetries
$$
\left.\left<\chi_{-}(\lambda), \chi_{-}(\lambda)\right>_{\pm} \right|_{\lambda=\lambda_1}, \ \ 
\left.(D_{\mu}+D_{\bar\mu})\left<\chi_{-}(\lambda), \chi_{-}(\lambda)\right>_{\pm} \right|_{\lambda=\lambda_1},
$$
and all symmetries of the type 
$$
(D_{\mu}+D_{\bar\mu})^k\Big(\left<\chi_{-}(\lambda), \chi_{-}(\lambda)\right> -\left<\chi_{+}(\lambda), \chi_{+}(\lambda)\right>_{\pm}\Big) \Big|_{\lambda=\lambda_1}, \ \ k\ge 0
$$
are $x$-periodic and isospectral. On the other hand, the symmetries
$$
(D_{\mu}+D_{\bar\mu})^k\Big(\left<\chi_{-}(\lambda), \chi_{-}(\lambda)\right> +\left<\chi_{+}(\lambda), \chi_{+}(\lambda)\right>\Big) \Big|_{\lambda=\lambda_1}, \ \ k\ge 1
$$ 
contain $x^2$ terms. Therefore we restrict our investigation to the symmetries depending on $x$ linearly:
\beq\label{chosen_sym}
\ba{l}
\left.(D_{\mu}+D_{\bar\mu})\left<\chi_{+}(\lambda), \chi_{-}(\lambda)\right>_{\pm} \right|_{\lambda=\lambda_1}= 
\left<\chi^{(1)}_{+}, \chi^{(0)}_{-}\right>_{\pm} , \\
\left.(D_{\mu}+D_{\bar\mu})^2\left<\chi_{+}(\lambda), \chi_{-}(\lambda)\right>_{\pm} \right|_{\lambda=\lambda_1}= 
\left<\chi^{(2)}_{+}, \chi^{(0)}_{-}\right>_{\pm} + 2 \left<\chi^{(1)}_{+}, \chi^{(1)}_{-}\right>_{\pm} ,\\
\left.(D_{\mu}+D_{\bar\mu})\left<\tilde\phi(\lambda),\tilde\phi(\lambda)\right>_{\pm} \right|_{\lambda=\lambda_1}= 
2\left<\tilde\phi^{(1)},\tilde\phi^{(0)}\right>_{\pm} ,
\ea
\eeq
since
\begin{equation}\label{aa}
\begin{split}
&\chi^{(1)}_{+}\circ \chi^{(0)}_{-}= ix~ \chi^{(0)}_{-}\circ \chi^{(0)}_{-}+\mbox{($x$-periodic part)}  ,\\
&\chi^{(2)}_{+}\circ \chi^{(0)}_{-}= 2ix~\chi^{(1)}_{-}\circ \chi^{(0)}_{-}+\mbox{($x$-periodic part)}  ,\\
&\chi^{(1)}_{+}\circ \chi^{(1)}_{-} = ix~\chi^{(1)}_{-}\circ \chi^{(0)}_{-}+\mbox{($x$-periodic part)}  ,\\
&\tilde\phi^{(1)}\circ \tilde\phi^{(0)}= ix~\tilde\phi^{(0)}\circ \tilde\phi^{(0)}+\mbox{($x$-periodic part)} .
\end{split}
\end{equation}

Since the explicit linear dependence on $x$ destroys the periodicity, we shall look now for suitable linear combinations of the symmetries \eqref{chosen_sym} that cancel this dependence. Evaluating explicitly the coefficients of $ix$ in \eqref{aa} (for simplicity at $t=0$) we have  

\beq
\ba{l}
\chi^{(0)}_{-}\circ \chi^{(0)}_{-} = \frac{4 \lambda_1^2\mu_1^2}{(1 - i \lambda_1 \sin(k\,x))^2} 
\begin{bmatrix} \bar q_2^2  \\ \bar q_1^2 \end{bmatrix}, \\[3mm]
\chi^{(1)}_{-}\circ \chi^{(0)}_{-}= \frac{2 \mu_1}{(1 - i \lambda_1 \sin(k\,x))^2} 
\left(\mu_1 [1 - i \lambda_1 \sin(k\,x)]\begin{bmatrix} r_1q_2 \\ r_2q_1 \end{bmatrix}+ [\lambda_1^2  + i \lambda \sin(k\,x) ]\begin{bmatrix} q_2^2  \\ q_1^2 \end{bmatrix} \right), \\[3mm]
\tilde\phi^{(0)}\circ \tilde\phi^{(0)} = \frac{\mu_1^2}{(1 - i \lambda_1 \sin(k\,x))^2} 
\begin{bmatrix} -\mu_1^2 - 2i\lambda_1\mu_1  \cos(k\,x) +\lambda_1^2 \cos^2(k\,x)     \\   \mu_1^2 - 2i\lambda_1\mu_1  \cos(k\,x) - \lambda_1^2 \cos^2(k\,x)   \end{bmatrix}.
\ea
\eeq
Consequently:
\begin{equation}\label{-}
\begin{split}
i\left[\left(\chi^{(0)}_{-}\circ \chi^{(0)}_{-}\right)_{1}-\overline{\left(\chi^{(0)}_{-}\circ \chi^{(0)}_{-}\right)_{2}}\right]&=-\frac{16i \lambda_1^2\mu_1^2}{(1 - i \lambda_1 \sin(k\,x))}, \\
i\left[\left(\chi^{(1)}_{-}\circ \chi^{(0)}_{-}\right)_{1}-\overline{\left(\chi^{(1)}_{-}\circ \chi^{(0)}_{-}\right)_{2}}\right]&=
-\frac{8i \mu_1\left[2 \mu_1^2- 1  + i \lambda_1 \sin(k\,x))    \right]}{(1 - i \lambda_1 \sin(k\,x))}, \\
i\left[\left(\tilde\phi^{(0)}\circ\tilde\phi^{(0)}\right)_{1}-\overline{\left(\tilde\phi_{(0)}\circ\tilde\phi^{(0)}\right)_{2}}\right]&= -\frac{2i\mu_1^2 (1 + i \lambda_1  \sin(k\,x)) }{(1 - i \lambda_1 \sin(k\,x))},
\end{split}
\end{equation}
and
\begin{equation}\label{+}
\begin{split}
\left(\chi^{(0)}_{-}\circ \chi^{(0)}_{-}\right)_{1}+\overline{\left(\chi^{(0)}_{-}\circ \chi^{(0)}_{-}\right)_{2}}&=\frac{16 \lambda_1^2\mu_1^3\, \cos(k\,x) }{(1 - i \lambda_1 \sin(k\,x))^2},\\
\left(\chi^{(1)}_{-}\circ \chi^{(0)}_{-}\right)_{1}+\overline{\left(\chi^{(1)}_{-}\circ \chi^{(0)}_{-}\right)_{2}}&= \frac{8 \mu_1^4  \cos(k\,x)}{(1 - i \lambda_1 \sin(k\,x))^2},\\
\left(\tilde\phi^{(0)}\circ\tilde\phi^{(0)}\right)_{1}+\overline{\left(\tilde\phi_{(0)}\circ\tilde\phi^{(0)}\right)_{2}}&= \frac{- 4 i\lambda_1 \mu_1^3 \cos(k\,x) }{(1 - i \lambda_1 \sin(k\,x))^2}.
\end{split}
\end{equation}

From the RHS of equations \eqref{-} it follows that, in order to cancel the linear dependence on $x$, we have to consider the following linear combination of all the three symmetries \eqref{chosen_sym}: 
\begin{equation}
\begin{split}  
  &Sym_1= 2 l_0^2\bigg[ m_0\left(\left<\chi^{(2)}_{+}, \chi^{(0)}_{-}\right>_+ + 2\left<\chi^{(1)}_{+}, \chi^{(1)}_{-}\right>_+\right) \\
  &- 4 \left<\chi^{(1)}_{+}, \chi^{(0)}_{-}\right>_+ - 16 \left<\phi^{(1)}_{+}, \phi^{(0)}_{-}\right>_+    \bigg],
\end{split}
\end{equation}
where $\mu_1=m_0$ and $\lambda_1= i l_o$. In this way, we obtain the following $x$-periodic symmetry
\beq\label{Sym1}
Sym_1(x,t)=k\frac{Num_1(x,t)}{\cal{D}(x,t)},
\eeq
where
\begin{equation}\label{Sym11}
\begin{split}
Num_1  &=
\big[ 192 i  k^4 \cos(k x) \sinh(\sigma t)- 80 i k^6 \cos(k x) \sinh(\sigma t)+ 8 i k^8 \cos(k x) \sinh(\sigma t) \\
  &+48 k^4 \sigma \cos(k x) \cosh(\sigma t)-8 k^6 \sigma \cos(k x) \cosh(\sigma t)\big] t+16 k^2 \cos(3 k x) \sinh(\sigma t)\\
  &-8 k^4 \cos(3 k x) \sinh(\sigma t)+ k^6 \cos(3 k x)\sinh(\sigma t)+24 k^4 \cos(k x) \sinh(\sigma t)\\
  &- k^6 \cos(k x)\sinh(\sigma t)+32 k^2 \sigma \sinh(2 \sigma t)-8 k^4 \sigma \sinh(2 \sigma t)-48 k^2 \cos(k x) \sinh(3 \sigma t) \\
  &+16 k^4 \cos(k x) \sinh(3 \sigma t)-64 k^2 \cos(k x) \sinh(\sigma t)-32 k \sigma \sin(2 k x) \sinh(2 \sigma t)\\
  &+8 k^3 \sigma \sin(2 k x) \sinh(2 \sigma t)- 64 i k \sin(2 k x)+i k^4 \sigma \cos(3 k x) \cosh(\sigma t)-i k^4 \sigma \cos(k x) \cosh(\sigma t) \\
  &+ 48 i k^3 \sin(2 k x)-8 i k^5 \sin(2 k x)- 4 i k^6 \cos(2 k x)- 64 i k^2 \cos(2 k x)- 48 i k^4 \cosh(2 \sigma t) \\
  &+ 32 i k^4 \cos(2 k x)+ 64 i k^2 \cosh(2 \sigma t)+ 8 i k^6 \cosh(2 \sigma t)- 64 i \sigma \cos(k x) \cosh(\sigma t) \\
  &- 16 i \sigma \cos(k x) \cosh(3 \sigma t)- 64 i k \sin(2 k x) \cosh(2 \sigma t)+ 48 i k^3 \sin(2 k x) \cosh(2 \sigma t)\\
  &- 8 i k^5 \sin(2 k x) \cosh(2 \sigma t)+ 16 i \sigma \cos(3 k x) \cosh(\sigma t)- 128 i k^2+ 48 i k^4- 4 i k^6\\
  &+ 16 i k^2 \sigma \cos(k x) \cosh(3 \sigma t)+ 40 i  k^2\sigma \cos(k x) \cosh(\sigma t)-8 i k^2 \sigma \cos(3 k x) \cosh(\sigma t),\\
  & \ \\
  \cal{D}(x,t)&=k  \, Den^2(x,t)= \\
    &=4\,\left[4\,k\cosh^2(\sigma\,t)+4\,\sigma\,\sin(k\,x)\,\cosh(\sigma\,t)+k(4-k^2)\sin^2(kx)\right],
\end{split}
\end{equation}
growing exponentially as $O(\exp(\sigma |t|))$ when $|t|\to \infty$.

From the RHS of equations \eqref{+} it follows that, in order to cancel the linear dependence on $x$, it is enough to consider the following linear combination of the first two symmetries in \eqref{chosen_sym}:
\begin{equation}
  Sym_2= 2 l_0^2\bigg( \left<\chi^{(2)}_{+}, \chi^{(0)}_{-}\right>_- + 2\left<\chi^{(1)}_{+}, \chi^{(1)}_{-}\right>_-
  - \frac{2m_0}{l_0^2} \left<\chi^{(1)}_{+}, \chi^{(0)}_{-}\right>_-  \bigg).
\end{equation}
Explicitly we have
\beq\label{Sym2}
Sym_2(x,t)=\frac{{Num}_2(x,t)}{{\cal D}(x,t)},
\eeq
where
\begin{equation}\label{Sym22}
\begin{split}  
  Num_2&=\big[ 512 k^2\sigma \sin(k x)\sinh(\sigma t)  + 256 i k\sigma \cosh(2 \sigma t) - 2176 i k^4 \sin(k x)\cosh(\sigma t)  \\
  &-416 k^4\sigma \sin(k x)\sinh(\sigma t)  +48 k^6\sigma \sin(k x)\sinh(\sigma t)  + 608 i k^6 \sin(k x)\cosh(\sigma t) \\
  &-48 i k^8 \sin(k x)\cosh(\sigma t) + 2048 i k^2 \sin(k x)\cosh(\sigma t) + 64 i k^5 \sigma+64 k^7 \sinh(2 \sigma t) \\
  &- 256 i  k^3\sigma \cosh(2 \sigma t)+ 192 i  k^3\sigma \cos(2 k x)- 32 i  k^5\sigma \cos(2 k x)- 256 i  k\sigma \cos(2 k x) \\
  &-512 i k^3 \sigma+ 64 i  k^5\sigma \cosh(2 \sigma t)+ 512 i k \sigma+512 k^3 \sinh(2 \sigma t)-384 k^5 \sinh(2 \sigma t)\big] t \\
  &-64  k\sigma  \cos(2 k x)\cosh(2 \sigma t)+16 k^3 \sigma  \cos(2 k x)\cosh(2 \sigma t)+128 k \sigma \\
  &-32 k^2 \sin(3 k x)\cosh(\sigma t)  -64  k\sigma \cos(2 k x)+128  k\sigma \cosh(2 \sigma t)+640 k^2  \sin(k x)\cosh(\sigma t) \\
  &+ 96 k^2  \sin(k x)\cosh(3 \sigma t)-112 k^4 \sin(k x) \cosh(\sigma t)+6 k^6  \sin(k x)\cosh(\sigma t)\\
  &+16 k^4 \sin(3 k x) \cosh(\sigma t)-2 k^6  \sin(3 k x)\cosh(\sigma t)-32 k^4 \sin(k x) \cosh(3 \sigma t)\\
  &+16  k^3\sigma \cos(2 k x)+ 128 i  k\sinh(2 \sigma t)- 128 i k^3 \sinh(2 \sigma t)- 128 i k \cos(2 k x) \sinh(2 \sigma t)\\
  &- 32 i \sigma \sin(3 k x) \sinh(\sigma t)+ 96 i k^3 \cos(2 k x) \sinh(2 \sigma t)- 16 i k^5 \cos(2 k x) \sinh(2 \sigma t)\\
  &+ 128 i \sigma \sin(k x) \sinh(\sigma t)+ 32 i \sigma \sin(k x) \sinh(3 \sigma t)- 2 i k^4 \sigma \sin(3 k x) \sinh(\sigma t)\\
  &+ 16 i k^2 \sigma \sin(3 k x) \sinh(\sigma t)- 80 i k^2\sigma \sin(k x) \sinh(\sigma t)+ 6 i k^4 \sigma \sin(k x) \sinh(\sigma t) \\
  &- 32 i k^2 \sigma \sin(k x) \sinh(3 \sigma t).
\end{split}
\end{equation}
Also this symmetry is $x$-periodic and grows exponentially as $O(\exp(\sigma |t|))$ when $|t|\to \infty$.\\

To simplify the above solutions, let us shift the AB \eqref{AB1} and the solutions \eqref{Sym1} and \eqref{Sym2} through the transformation $x\rightarrow x-L/4$, obtaining for the AB an even function in $x$:
\beq\label{AB2}
u(x,t)=\frac{(2 k^2-\sigma^2) \cosh(\sigma t)+ i k^2 \sigma\sinh(\sigma t) +  k\sigma\cos(k x) }
{k(2\cosh(\sigma t) k-\sigma\ cos(kx))}.
\eeq
It follows that the linearized equation \eqref{linearizedNLS} becomes invariant under parity transformation: if $w(x,t)$ is solution, also $w(-x,t)$ is solution, as well as its even and odd parts.

Denote by $\widehat{Sym}_1(x,t)=\frac{1}{2} \left( Sym_1(x,t) -Sym_1(-x,t)  \right)$ the odd part of $Sym_1(x,t)$. Then

\beq\label{Sym1bis}
\widehat{Sym}_1(x,t)=k\frac{\widehat{Num_1}(x,t)}{\cal D(x,t)}
\eeq
\begin{equation}
\begin{split}
  &\widehat{Num_1}(x,t)=
  \bigg[\big[48 \sigma k^4-8 \sigma k^6\big] \cosh(\sigma t)+\big[ 192 i k^4+ 8 i  k^8- 80 i k^6\big] \sinh(\sigma t)\bigg] t \sin(k x) + \\
  &+ \bigg[\big[- 16 i  \sigma+ 16 i \sigma k^2 \big] \cosh(3 \sigma t) +
  \big[- 64 i  \sigma+ 40 i \sigma k^2-i k^4 \sigma \big] \cosh(\sigma t) + \\
  &+ \big[16 k^4-48 k^2 \big] \sinh(3 \sigma t)+ \big[-64 k^2-k^6+24 k^4\big] \sinh(\sigma t)\bigg] \sin(k x)+\\
  &+
  \bigg[ \big[-48 i  k^3+ 64 i k+ 8 i k^5\big] \cosh(2 \sigma t)+\big[32 \sigma k-8 \sigma k^3 \big] \sinh(2 \sigma t)+\\
  &+ \big[8 i k^5- 48 i  k^3+ 64 i  k\big] \bigg] \sin(2 k x)+ \\ &
  +\bigg[\big[8 i \sigma k^2-i \sigma k^4- 16 i \sigma \big] \cosh(\sigma t) +\big[8 k^4 -16 k^2 -k^6\big] \sinh(\sigma t)\bigg] \sin(3 k x),
\end{split}
\end{equation}
$$
\cal{D}(x,t)=4\,\left[4\,k\cosh^2(\sigma\,t)-4\,\sigma\,\cosh(\sigma\,t)\,\cos(k\,x)+k(4-k^2)\cos^2(kx)\right].
$$

Under the transformation $x\rightarrow x-L/4$ the solution  $Sym_2$ becomes even in $x$, and reads
\beq\label{Sym2bis}
Sym_2(x,t)= \frac{Num_2(x,t)}{\cal D(x,t)}
\eeq
\begin{equation}
\begin{split}
  &Num_2(x,t)=\\
  & \bigg[ 256 i \sigma k- 192 i  k^3 \sigma+ 32 i k^5 \sigma \bigg]\  t \cos(2 k x)+\\
  &+\bigg[\big[ 2176 i  k^4+ 48 i k^8- 2048 i k^2- 608 i k^6\big] \cosh(\sigma t)+ \\
  &+\big[ 416 \sigma k^4-48 \sigma k^6-512 \sigma k^2\big] \sinh(\sigma t)
  \bigg]\  t \cos(k x) -\\
  &- 64 k \bigg[ \big[6 k^4-8 k^2-k^6 \big]\sinh(2 \sigma t)+\big[ 4 i \sigma k^2-i k^4 \sigma- 4 i \sigma\big] \cosh(2 \sigma t)+ \\
  & +\big[-i k^4 \sigma + 8 i \sigma k^2- 8 i \sigma\big] \bigg] \ t + \\
  &+ \bigg[\big[-2 k^6-32 k^2+16 k^4\big] \cosh(\sigma t)+ \big[- 2 i  \sigma k^4+ 16 i  \sigma k^2- 32 i \sigma\big] \sinh(\sigma t) \bigg] \cos(3 k x) +\\
   &+ \bigg[\big[64 \sigma k-16 \sigma k^3\big] \cosh(2 \sigma t)+ \big[- 96 i k^3+ 128 i k+ 16 i k^5\big] \sinh(2 \sigma t)+ \\
  &+\big[64 \sigma k-16 \sigma k^3\big] \bigg]\ \cos(2 k x) +\\
  &+\bigg[\big[-96 k^2+32 k^4\big] \cosh(3 \sigma t)+ \big[-640 k^2+112 k^4-6 k^6\big] \cosh(\sigma t) + \\
  &+\big[ 32 i \sigma k^2- 32 i \sigma\big] \sinh(3 \sigma t)+
  \big[- 6 i \sigma k^4+ 80 i  \sigma k^2- 128 i \sigma\big] \sinh(\sigma t)\bigg] \cos(k x)-\\ 
  &- 64 k \bigg[  \big[2 i k^2-2 i\big] \sinh(2 \sigma t)-2 \sigma \cosh(2 \sigma t)-2 \sigma \bigg].
\end{split}
\end{equation}

Therefore we have constructed two examples of $x$-periodic symmetries of NLS growing exponentially when $|t|\to\infty$. We conclude that the Akhmediev breather is linearly unstable with respect to generic periodic perturbations if $N=M=1$, proving rigorously what was rather evident from the first qualitative argument of \S 2. This result can be generalized to the case of the $AB_M$ solution, for $M\le N$. We conclude that the $AB_M$, $M\in\NN^+$ is linearly unstable with respect to generic periodic perturbations.

\begin{remark}
Formulas (\ref{Sym1bis}), (\ref{Sym2bis}) for the symmetries $\widehat{Sym}_1(x,t)$ and $Sym_2(x,t)$ have been calculated using the computer algebra Maple~15 system. Using Maple~15 it was also checked directly that these expressions satisfy the NLS equation \eqref{linearizedNLS} linearized about the Akhmediev breather (\ref{AB2}).
\end{remark}

\section{Nonlinear instability of the AB initial condition, and an  AW recurrence of ABs}  

The linear instability of the AB proven in the previous section does not allow one to say anything about the nonlinear evolution of this instability. This section is devoted to this goal.

As we know, the Akhmediev breather
\beq\label{Akhm1}
\ba{l}
{\cal A}(x,t;\phi,X,T)=e^{2it}\frac{\cosh[\sigma (t-T)+2i\phi ]+\sin\phi \cos[k (x-X)]}{\cosh[\sigma (t-T)]-\sin\phi \cos[k (x-X)]},\\
\sigma=k\sqrt{4-k^2}=2\sin(2\phi), \ \ k=k_1=\frac{2\pi}{L}=2\cos\phi,
\\ \phi=\arccos\left(\frac{k}{2}\right)=\arccos\left(\frac{\pi}{L}\right),
\ea
\eeq
describes the instability of the mode $k=k_1$ in the periodic setting, where $X$ and $T$ are respectively the position and time of appearance of the coherent structure.

Expanding this solution in Fourier series:
\beq\label{akhmed}
{\cal A}(x,t;\phi,X,T)=\sum_{n\in\ZZ}C_n(t,X,T)e^{i k_n x}, \ \ k_n=n k_1,
\eeq
its Fourier coefficients have a simple analytic expression, obtained via standard contour integration:
\beq
\ba{l}
C_n(t,X,T)=\frac{1}{L}\int\limits_0^L e^{-i k_n x}{\cal A}(x,t;\phi,X,T)=\\
e^{2it}e^{-i k_n X}\left(-\delta_{n0}+\frac{A(t-T,\phi)+B(t-T,\phi)}{\sqrt{B^2(t-T,\phi)-1}}\left(B(t-T,\phi)-\sqrt{B^2(t-T,\phi)-1}\right)^{|n|}\right),
\ea
\eeq
where
\beq
A(t-T,\phi)=\frac{\cosh[\sigma(t-T)+2i\phi]}{\sin\phi}, \ B(t-T,\phi)=\frac{\cosh[\sigma(t-T)]}{\sin\phi}.
\eeq
If we choose $T$ positive and sufficiently large to have
\beq
\ba{l}
\delta:=\exp(-\sigma T)\ll 1 ,
\ea
\eeq
at the initial time $t=0$ the AB (\ref{akhmed}) is itself an $O(\delta)$ perturbation of the background. Since
\beq
C_0=\exp(-2 i \phi)(1+O(\delta^2)),
\eeq
to be consistent with the notation used in \eqref{eq:nls_cauchy1}, in which the initial background is $1$, we normalize the Fourier coefficients as follows:
\beq
c_n:=C_n/C_0, \ \ n\in\ZZ,
\eeq
obtaining, in particular, $c_0=1$ and 
\beq
c_{\pm 1}(0,X,T)=\delta e^{2 i t\mp i k_1 X}\sin(\phi)(1+e^{2 i \phi})\left(1+O(\delta^2)\right).
\eeq
Therefore the $\alpha,\beta$ coefficients \eqref{def_alpha_beta} of the Akhmediev initial condition read 
\beq
\ba{l}
\alpha_{AB}=e^{-i\phi}\overline{c_1(0,T,X)}-e^{i\phi}c_{-1}(0,T,X)= -2i\delta e^{i k_1 X}\sin^2(2\phi)(1+O(\delta^2)) , \\
\beta_{AB}=e^{i\phi}\overline{c_{-1}(0,T,X)}-e^{-i\phi}c_1(0,T,X)=O(\delta^3).
\ea
\eeq
It follows that a small periodic perturbation of the AB initial condition must be taken, without loss of generality, in the form
\beq\label{Cauchy_1}
\ba{l}
u(x,0)=e^{2 i\phi}{\cal A}(x,0,\phi,X,T)+\eps v(x), \ \ v(x+L)=v(x), \\
v(x)=\sum\limits_{n\ge 1}\left(v_n e^{i k_n x}+v_{-n} e^{-i k_n x}\right), \ \ v_n=O(1),  
\ea
\eeq
where
\beq
\delta^3\ll \eps \ll \delta .
\eeq
Then the $\alpha$ and $\beta$ coefficients \eqref{def_alpha_beta} of this initial condition read: 
\beq\label{alpha_beta_pert}
\ba{l}
\tilde\alpha=-2i\delta e^{i k_1 X}\sin^2(2\phi) +\mbox{ max}\left(O(\eps),O(\delta^2)\right), \\
\tilde\beta=\eps\left(e^{i\phi}\overline{v_{-1}}-e^{-i\phi}v_1\right) + \mbox{ max}\left(O(\delta^3),O(\eps^2) \right).
\ea
\eeq
Applying formulas \eqref{unif_sol_Cauchy_1}-\eqref{parameters_1n} and the theory developed in \cite{GS1,GS2,GS3}, it follows that the evolution of the initially perturbed AB (\ref{Cauchy_1}) is described, to leading order, by the following recurrence of ABs
\begin{align}\label{unif_sol_Cauchy_2}
u(x,t)&=\sum\limits_{m=0}^n e^{i\rho^{(m)}}{\cal A}\Big(x,t;\phi,x^{(m)},t^{(m)} \Big)- \\
 &-\frac{1-e^{4in\phi}}{1-e^{4i\phi}} e^{2it}, \ \ x\in [0,L], \nonumber
\end{align}
where the parameters $x^{(m)},~ t^{(m)},~\rho^{(m)},~m\ge 0$, are defined as:
\beq\label{parameters_2n}
\ba{l}
x^{(m)}=x^{(1)}+(m-1)\Delta X, \ \ t^{(m)}=t^{(1)} + (m-1)\Delta T, \\
x^{(1)}=\frac{\arg\tilde\alpha}{k_1} +\frac{L}{4}=X, \ \Delta X =\frac{\arg(\tilde\alpha\tilde\beta)}{k_1}=X+\frac{\arg(\tilde\beta)}{k_1}-\frac{L}{4}, \ (\!\!\!\!\!\mod L),\\
t^{(1)}\equiv \frac{1}{\sigma_1}\log\left(\frac{\sigma^2_1}{2|\tilde\alpha|} \right)=T, \ \
\Delta T = \frac{1}{\sigma_1} \log\left(\frac{\sigma^4_1}{4|\tilde\alpha\tilde\beta|}\right)=T+
\frac{1}{\sigma_1} \log\left(\frac{\sigma^2_1}{2|\tilde\beta|}\right), \,  \\
\rho^{(m)}=2\phi_1+(m-1)4\phi_1 ,
\ea
\eeq
and $\tilde\alpha$ and $\tilde\beta$ are defined in \eqref{alpha_beta_pert} (see Figures \ref{a1}).

As in the case of the initial perturbation of the pure background, the solution (\ref{unif_sol_Cauchy_2}), (\ref{parameters_2n}) shows an exact recurrence of AWs described, to leading order, by the AB, whose parameters change at each appearance according to (\ref{parameters_2n}). It is another good example of a FPUT type recurrence without thermalisation. $x^{(1)}=X$ and $t^{(1)}=T$ are respectively the position and the time of the first appearance, coinciding to leading order with the appearance of the AB as if it were not perturbed; $\Delta X$ is the $x$-shift of the position of the AW between two consecutive appearances, and $\Delta T$ is the recurrence time (the time between two consecutive appearances). 

While, in the previous section, we have proven the linear instability of the AB with respect to initial perturbations, the theory developed in \cite{GS1,GS2,GS3} allows one to prove, through formulas (\ref{unif_sol_Cauchy_2}), (\ref{parameters_2n}), the nonlinear instability  of the AB solution with respect to initial perturbations, providing the proper analytic model to study quantitatively and in terms of elementary functions how this instability generates the above FPUT recurrence. We remark that the qualitative features of this nonlinear instability are quite predictable from the second qualitative argument of \S 2.  

If we compare the $O(\sigma^{-1}|\log(\eps^2)|)$ time recurrence generated by a generic $O(\eps)$ perturbation of the unstable background $\exp(2it)$ (see \eqref{parameters_1n}), with the $O(\sigma^{-1}|\log(\eps\delta))|$ time recurrence obtained perturbing the AB with a generic $O(\eps)$ perturbation (see \eqref{parameters_2n}), we observe that
\beq
O\left(\sigma^{-1}|\log(\eps\delta)| \right)   <   O\left(\sigma^{-1}|\log(\eps^2)|\right),
\eeq
inferring that, for a generic $O(\eps)$ perturbation, the AB is more unstable than the background solution (see Figures \ref{a1}).

To complete the analysis, one should also consider the case in which the first appearance time $T$ of the AB initial condition is closer to the origin, so that $\exp(-\sigma T)= 0(1)$. As we shall see in a subsequent paper, an $0(\eps)$ perturbation of an AB initial condition of this type generates $0(\sqrt{\eps})$ gaps, implying a time recurrence of $O\left(\sigma^{-1}|\log(\eps)|\right)$, such that
\beq
O\left(\sigma^{-1}|\log(\eps)|\right)<O\left(\sigma^{-1}|\log(\eps\delta)| \right)   <   O\left(\sigma^{-1}|\log(\eps^2)|\right).
\eeq
Therefore, for a generic $O(\eps)$ perturbation, the AB is more unstable than the background, and it becomes more and more unstable as $T\to 0$ (see Figures \ref{a1}).

\section{Conclusions, open problems, and remarks}

The AB is unstable with respect to perturbations of the focusing NLS equation, as it was shown in real and numerical experiments \cite{Kimmoun,Soto}, and as it was analytically proven in \cite{CGS}, where the NLS perturbation theory for AWs was constructed and applied to the case of the NLS perturbed by linear loss or gain terms, and in \cite{CS1}, where such a theory was applied to the complex Ginzburg-Landau \cite{Newell_Whitehead} and Lugiato Lefever \cite{LL} equations, viewed as perturbations of focusing NLS.

It is also unstable with respect to perturbations due to the discrete scheme used in the numerical experiments \cite{AblowHerbst,AblowSchobHerbst,GS4}.  In particular, in \cite{GS4} it was recently shown that, when the initial condition is given by the unperturbed AB, in the simplest case of $M=N=1$, and in the situation in which the round-off errors are negligible with respect to the perturbations due to the discrete scheme used in the numerical experiments, the Split-Step Fourier Method (SSFM) \cite{JR}, the numerical output exhibits again an exact recurrence of AW described, at each appearance, by the AB. In addition, a remarkable empirical formula connecting the recurrence time with the number of time steps used in the SSFM was discovered.

\begin{figure}[H]
\begin{center}
  \includegraphics[width=4.5cm]{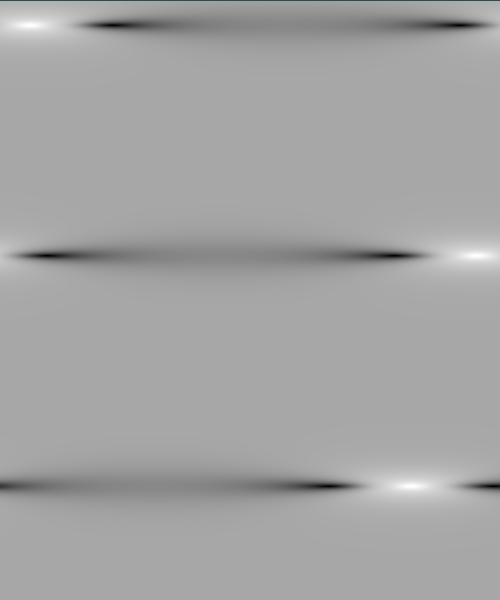}~~~~
  \includegraphics[width=4.5cm]{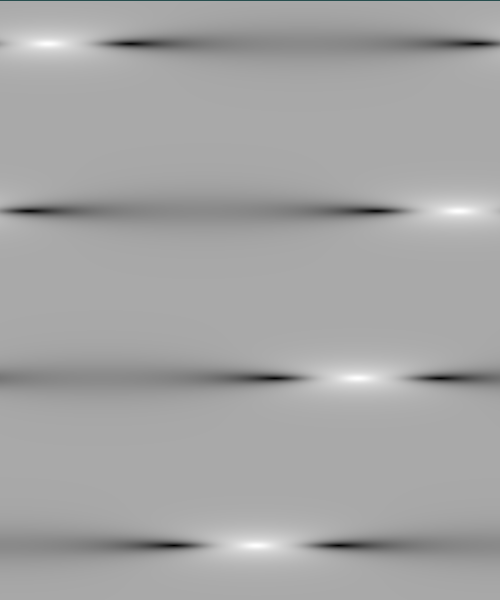}\\[3mm]
  \includegraphics[width=4.5cm]{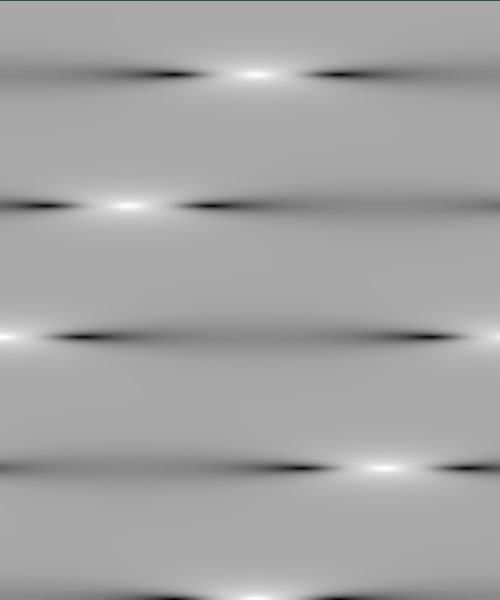}~~~~
  \includegraphics[width=4.5cm]{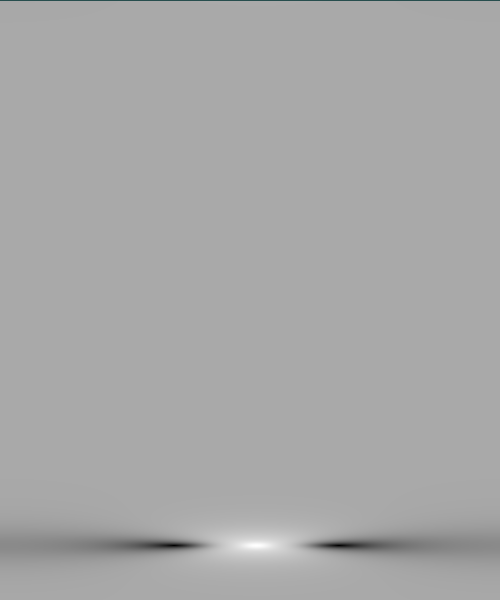}
  \caption{\label{a1} Density plots of $|u(x,t)|$ obtained through the numerical integration of NLS using the SSFM  with quadrupole precision, for $|x|\le L/2,~L=6$, and $t\in[0,30]$. In the first three pictures we use the same perturbation: $\eps=10^{-4}$, $v_1=0.1-i~ 0.5,~v_{-1}=-0.1+i~ 0.1$. We have the following initial conditions. Top left: perturbed background. Top right: perturbed AB with $T=2.7$ (where $T$ is AB appearance time parameter in \eqref{eq:akh1}), corresponding to $\delta=0.81\cdot 10^{-2}\gg \eps$; this numerical experiment is in good quantitative agreement with the analytic formulas \eqref{alpha_beta_pert}-\eqref{parameters_2n}. Bottom left: perturbed AB with $T=0$. Bottom right: unperturbed AB with $T=2.7$. The recurrence times of the first three experiments confirm that the AB is more unstable than the background, and that it becomes more and more unstable as $T\to 0$. Comparing the top right and bottom right pictures we confirm that the first appearance is essentially $T$ in both cases. We also remark that, if we had used a sufficiently longer time interval of integration, then we would have observed a recurrence of ABs also in the bottom right picture, corresponding to the unperturbed AB initial condition, as shown in \cite{GS4}.}
\end{center}
\end{figure}

In this paper we concentrated instead on the linear and nonlinear instability properties of the AB within the NLS dynamics, obtaining the following results. \\
1) We constructed two explicit examples of $x$-periodic symmetries of the NLS equation growing exponentially in $t$, when $|t|\to\infty$, thus proving the linear instability of the AB, even in the so-called case of saturation of the instability, in contrast to what is commonly believed in the current literature. This result can be generalized to the case of the $M$-soliton solution of Akhmediev type. \\
2) We studied the nonlinear instability of the AB, showing that a small periodic perturbation of the AB initial condition evolves into a recurrence of ABs of FPUT type, whose parameters are different at each appearance, thus describing in terms of elementary functions the nonlinear instability of the AB solution (see Figures \ref{a1}).\\
3) At last we showed that, with respect to the same $O(\eps)$ initial perturbation, the AB solution is more unstable than the background solution of NLS, and it is more and more unstable as $T\to 0$, where $T$ is its appearance time parameter (see Figures \ref{a1}).

In future works on this subject we plan: i) to provide a regular procedure for constructing the squared eigenfunction decomposition, the basic tool in the theory of perturbations, for non-generic situations like the one discussed in this paper; ii) to use this result for constructing the analytic description of the FPUT recurrence of ABs in the case in which its first appearance time $T$ is close to the origin. 

We end the paper with an important remark: \textbf{inspite of its linear and nonlinear instability properties, the AB is relevant in nature, since its instability leads to a FPUT recurrence of ABs. This result suitably generalizes to the case $M>1$}.

\ \\
\ \\

\ \\
\ \\

\end{document}